\documentclass{elsart}

\usepackage{amssymb,amsmath,mathrsfs,eufrak,mathabx}

\newtheorem{exm}{Example}[section]

\newcommand{\A}{\bf{A} \it}

\newcommand{\ar}{\rightarrow}
\newcommand{\la}[1]{\it {#1}\rm}

\newcommand{\mL}{\mathcal{L}}

\newcommand{\mA}{\mathcal{A}}

\newcommand{\mV}{\mathcal{V}}

\newcommand{\mS}{\mathcal{S}}

\newcommand{\bF}{{\bf F}}

\newcommand{\mF}{\mathcal{F}}

\newcommand{\mI}{\mathcal{I}}

\newcommand{\lm}[1]{\begin{lem} #1 \end{lem}}
\newcommand{\te}[1]{\begin{thm} #1 \end{thm}}
\newcommand{\ex}[1]{\begin{exm} #1 \end{exm}}
\newcommand{\co}[1]{\begin{cor} #1 \end{cor}}
\newcommand{\rk}[1]{\begin{rem} #1 \end{rem}}

\newcommand{\bA}{{\bf A}}
\newcommand{\bB}{{\bf B}}

\begin{document}

\begin{frontmatter}



\title{ Algebraic Logic, I
\\ \small{Quantifier Theories and Completeness Theorems}
}


{\large Zhaohua Luo}
\author{}


\ead{zluo@algebraic.net}
\ead[url]{http://www.algebraic.net/cag/alglog.html}

\begin{abstract}

 Algebraic logic studies algebraic theories related to proposition and first-order logic. A new algebraic approach to first-order logic is sketched in this paper. We introduce the notion of a quantifier theory, which is a functor from the category of a monad  of sets to the category of Boolean algebras, together with a uniquely determined system of quantifiers. A striking feature of this approach is that Cayley's Completeness Theorem and G\"{o}del's Completeness Theorem can be stated and proved  in a much simpler fashion for quantifier theories. Both theorems are due to Halmos for polyadic algebras. We also present a simple transparent treatment of ultraproducts of models of a quantifier theory.

\end{abstract}

\end{frontmatter}

\bf{Content}\rm
\\ 0. Introduction.
\\ 1. Properties of Binding Theories.
\\ 2. Functional Theories.
\\ 3. Cayley's Completeness Theorem.
\\ 4. G\"{o}del's Completeness Theorem.
\\ 5. Quantifier Theories with Equality.
\\ 6. Ultraproducts of Models.
\\ 7. Polyadic Theories.

\section*{Introduction}

Algebraic logic studies algebraic theories related to proposition and first-order logic. A new algebraic approach to first-order logic is sketched in this paper. We introduce the notion of a quantifier theory and prove Cayley's Completeness Theorem and G\"{o}del's Completeness Theorem for quantifier theories. Both theorems are due to Halmos for polyadic algebras. We also present a simple transparent treatment of ultraproducts of models of a quantifier theory. This approach to algebraic logic is based on the theory of clones (see \cite{luo:1}-\cite{luo:4}).

It is well known that Boolean algebras algebrazies proposition logic, and polyadic algebras algebraizes first-order logic. In literature polyadic algebras are usually defined as a substitution Boolean algebra over a \la{fixed set of variables without terms} (cf. \cite{halmos:1}). In order to prove G\"{o}del's completeness theorem for polyadic algebras one needs to add constants to a polyadic algebra as new closed terms. Halmos's original approach  to the theory of constants and terms for polyadic algebra are quite involved. To overcome these conceptual difficulties we introduce the notion of a \la{quantifier theory}, which is a functor from the Kleisli category of a monad of sets to the categories of Boolean algebras, equipped with a \la{binding system} of \la{quantifiers}. This is a very natural approach to first-order logic as the semantic or syntax of quantifier logic provide concrete or abstract quantifier theories respectively (see Section \ref{section:function} and the last part of this introduction).

By a \la{Boolean algebra} $B$  we mean a complemented distributive lattice, which may be viewed as an algebra $(B, \wedge, \vee, \neg, 0, 1)$ with two binary operations $\wedge, \vee$, a unary operation $\neg$, and two distinguished elements $0, 1$.  Alternatively, a Boolean algebra is an algebra $(B, \wedge, \neg)$ such that $(B, \wedge)$ is a commutative semigroup and $p \wedge (\neg q) = r \wedge (\neg r)$ iff $p \wedge q = p$ for any $p, q, r \in B$ (cf. \cite{bell:1} \cite{ros}).  A Boolean algebra is \la{nontrivial} if it has at lease two distinct elements (i.e., $0 \ne 1$). A nonempty subset $I$ of $B$ is \la{consistent} if  $p_1 \wedge ... \wedge p_n \ne 0$ for any nonempty finite subset $\{p_1, .., p_n\}$ of $I$. An \la{ultrafilter} of $B$ is a maximal consistent set. A \la{filter} of $B$ is an intersection of ultrafilters (see  Section \ref{sec:com}). Denote by $2 = \{0, 1\}$ the smallest nontrivial Boolean algebra.

\bf{Notation.} \rm Let $A, B$ be arbitrary  sets.
\\ 1. If $a, b \in A$ denote by $[b/a]: A \ar A$ the map sending $a$ to $b$ and  other element of $A$ to itself.
\\ 2. Suppose $\sigma: A \ar B$ is a map. If  $a \in A$ and $b \in B$,  we denote by $\sigma^{b/a}: A \ar B$ the map sending $a$ to $b$ and other element $c$ of $A$ to $\sigma(c)$.
\\ 3. Suppose $\sigma: A \ar B$ is a map. If $U \subseteq A$, $V \subseteq B$ and $\pi: U \ar V$ is a map, let $\sigma^{\pi}: A \ar B$ be the map such that $\sigma^{\pi}(a) = \pi(a)$ if $a \in U$ and $\sigma^{\pi}(a) = \sigma(a)$ otherwise.
\\ 4. $|A|$ denotes the cardinality of $A$.
\\ 5. If $A$ is a subset of $B$ denote by $\kappa_X$ (or simply $\kappa$) the inclusion map from $A$ to $B$.

Let $\mS$ be a nonempty collection of sets which contains at least one infinite set. Let $X, Y, Z, ... $ be any sets in $\mS$, viewed as \la{sets of variables}.

Let $\mV$ be a variety in the sense of universal algebra.

A  \la{binding theory $\bA$ (over $\mS$)} of $\mV$-algebras consists of
  \\ T1. a set $\bA^*(X)$ and  a map $\epsilon_X: X \ar \bA^*(X)$ for each set $X$ ( we often simply write $x$ for $\epsilon_X(x)$);
 \\ T2. an algebra $\bA_*(X)$ in $\mV$ for each set $X$;
  \\ T3. an element $a\sigma \in \bA^*(Y)$ for each nonempty set $X$, each element $a \in \bA^*(X)$ and each map $\sigma: X \ar \bA^*(Y)$;
  \\ T4. an element $p\sigma \in \bA_*(Y)$  for each nonempty set $X$, each element $p \in \bA_*(X)$ and each map $\sigma: X \ar \bA^*(Y)$;
 \\ T5. an element $\forall x.p \in \bA_*(X)$ for each nonempty set $X$, each variable $x \in X$ and each element $p \in \bA_*(X)$;

For any set $X$ let $\bA(X) = \bA_*(X) \cup \bA^*(X)$. If $X$ is nonempty and $\sigma: X \ar \bA^*(Y)$ is a map let $\sigma^*$ be the map sending $a \in \bA^*(X)$ to $a\sigma$ and let $\sigma_*$ be the map sending $p \in \bA_*(X)$ to $p\sigma$.  Denote by $\sigma*$$: \bA(X) \ar \bA(Y)$ the map sending $t \in \bA(X)$ to $t\sigma$.

We assume that a binding theory $\bA$ satisfies the following conditions for any nonempty $X$, $x \in X$, $t \in \bA(X)$, $\sigma: X \ar \bA^*(Y)$ and $p \in \bA_*(X)$:
\\ P1. $t\epsilon_X = t$;
\\ P2. $(t\sigma)\tau = t(\sigma\tau)$ if $Y$ is nonempty, $\tau: Y \ar \bA^*(Z)$ is a map, and $\sigma\tau$ is the map sending $x$ to  $\sigma(x) \tau$;
\\ P3. $x\sigma = \sigma(x)$ for any $x \in X$;
\\ P4. $\sigma_*: \bA_*(X) \ar \bA_*(Y)$ is a homomorphism of algebras in $\mA$.
\\ P5. $(\forall x.p)\sigma = \forall y.(p\sigma^{y/x})$ if $Y$ is nonempty, $\sigma[z/y] = \sigma$ for some $y, z \in Y$ such that $y \ne z$.

\rk{(a) If we omit T5 and F5 then a system $\bA$ thus defined is called a substitution theory of $\mV$-algebras. A function $\forall$ defined by T5 on a substitution theory $\bA$ satisfying P5 is called a binding system on $\bA$.
\\ (b) A system $\bA^*$ consisting of all $\bA^*(X)$ and $a\sigma$ satisfying P1-P3 is called a clone over $\mS$, and $\bA$ is called a theory over clone $\bA^*$ (see \cite{luo:2}).
}

 \rk{Suppose $\bA$ is a (substitution or binding) theory.
  \\ (a) We say $\bA$ is faithful if $\emptyset \in \mS$, $\bA^*(\emptyset)$ is nonempty, and for any nonempty set $X$ and $t, s \in A(X)$ we have $t = s$ iff $t\sigma = s\sigma$ for any $\sigma: X \ar \bA^*(\emptyset)$.
  \\ (b) We say $\bA$ is a global theory if $\mS$ is the category of sets.
}

Let $\bA$ be a substitution or binding theory. Suppose $X$ is a nonempty set and $t \in \bA(X)$. A subset $U$ of $X$ is called a \la{support} for $t$ if we have $t\sigma = t\tau$ for any $\sigma, \tau: X \ar \bA^*(Y)$ with $\sigma|_U =  \tau|_U$. Denote by $\A(X)_U$ the set of elements of $\bA(X)$ with $U$ as a support.  We say $t$ is \la{independent of} $U$ if $X\setminus U$ is a support for $t$.   We say $t$ is \la{closed} if $\emptyset$ is a support for $t$. By definition we assume any element in $\bA(\emptyset)$ is closed (if $\emptyset \in \mS$). A theory $\bA$  is \la{locally finite} if each element of $\bA(X)$ has a finite support.

Suppose $\bA$ is a substitution theory over $\mS$ and $\bB$ is a substitution theory over another collection $\mS'$ of sets such that $\mS \subseteq \mS'$. A \la{morphism} $\phi = (\phi_*, \phi^*): \bA \ar \bB$ consists of a map $\phi_X^*: \bA^*(X) \ar \bB^*(X)$  and a homomorphism $\phi_{X*}: \bA_*(X) \ar \bB_*(X)$ of algebras for each $X \in \mS$ such that for any $\sigma: X \ar \bA^*(Y)$:
\\ N1. $\phi_*(p)(\sigma\phi^*) = \phi_*(p\sigma) $ where $(\sigma\phi^*)(x) = \phi^*(\sigma(x))$;
\\ N2. $\phi^*(a)(\sigma\phi^*) = \phi^*(a\sigma) $;
\\ N3. $\phi^*(x) = x$.
\\ Here for simplicity we write $\phi^*$ for $\phi_X^*$ and $\phi_*$ for $\phi_{X*}$.
\\ If $\bA$ and $\bB$ are binding theories we also require that:
\\ N4. $\phi_*(\forall x.p) = \forall x.(\phi_*(p))$.
\\ A morphism $\phi: \bA \ar \bB$ is called an \la{embedding} if  $\phi_{X^*}$ and $\phi_X^*$ are injective for any $X$. We define the notion of a \la{subtheory of $\bA$} in an obvious way.

 A \la{quantifier theory} is a binding theory $\bA$ of Boolean algebra satisfying  the following conditions for any nonempty set $X$,  $x \in X$ and $p, q \in \bA_*(X)$:
\\ Q1. $\forall x. (p \wedge q) = \forall x. p \wedge \forall x. q$;
\\ Q2. $\forall x. p \leq  p$;
\\ Q3. $\forall x.p = p$ if $p[y/x] = p$ for some $y \in X$ such that $x \ne y$.
 \\ A quantifier theory over a set $\{X\}$ is called a \la{quantifier algebra over $X$}. A quantifier theory is \la{nontrivial} if $\bA_*(X)$ has a non-closed element for some nonempty set $X$. Unless otherwise stated all the quantifier theories considered below are nontrivial.

A \la{quantifier model} is a quantifier theory  $\bA$ over $\mS$ with $\emptyset \in \mS$ satisfying the following conditions:
 \\ M1. $\bA^*(\emptyset)$ is nonempty and $\bA_*(\emptyset)$ is a nontrivial Boolean algebra.
  \\ M2. For any $x \in X$, $p \in \bA_*(X)$, and $\sigma: X \ar \bA^*(\emptyset)$ we have $(\forall x.p)\sigma = \bigwedge_{d \in \bA^*(\emptyset)} p\sigma^{d/x}$.
  \\ We say $\bA$ is a \la{$2$-model} if  $\bA_*(\emptyset) = 2 = \{0, 1\}$.
 \\ Suppose $\bA$ is a quantifier theory over $\mS$. A \la{modification} (resp. model) of $\bA$ is a quantifier theory (resp. quantifier model) $\bB$ over $\mS \cup \{\emptyset\}$ such that $\bA|_{\mS\setminus \{\emptyset\}} = \bB|_{\mS\setminus \{\emptyset\}}$.

\rk{Suppose $\bA$ is a quantifier theory over $\mS$ and $\bB$ is a quantifier model over collection of sets containing $\mS$ and $\emptyset$. Any morphism $\phi$ from $\bA$ to $\bB$ induces a quantifier model $\bA(\phi)$ of $\bA$ with $\bA(\phi)_*(\emptyset) = \bB_*(\emptyset)$ and $\bA(\phi)^*(\emptyset) = \bB^*(\emptyset)$.}

  \rk{Suppose $\bA$ is a quantifier theory.
  \\ (a) Any nonempty set $Z \in \mS$ determines a modification $\bA[Z]$ of $\bA$ such that $\bA[Z]_*(\emptyset) = \bA_*(Z)$ and $\bA[Z]^*(\emptyset) = \bA^*(Z)$, with the given $t\sigma$ for any $t \in \bA(X)$ and $\sigma: X \ar \bA[Z]^*(\emptyset) = \bA(Z)$. The quantifier theory $\bA[Z]$ is called the modification of $\bA$ by $Z$.
  \\ (b) If $I$ is a filter of $\bA_*(\emptyset)$ and $\bA_*(\emptyset)/I$ is the quotient algebra of Boolean algebra $\bA_*(\emptyset)$ module $I$, we denote by $\bA/I$ the modification of $\bA$ with  $\bA_*(\emptyset) = \bA_*(\emptyset)/I$ and $\bA/I^*(\emptyset) = \bA^*(\emptyset)$.
   }

 \te{\label{te:cay11} (Cayley's Completeness Theorem For Quantifier Theories) Suppose $\bA$ is a locally finite quantifier theory and $Z \in \mS$ is an infinite set. Then the modification $\bA[Z]$ of $\bA$ by $Z$ is a faithful quantifier model of $\bA$.  }

 The following important theorem is a consequence of Cayley's completeness Theorem (see Theorem \ref{te:base-extension}).

\te{\label{base} (a) For any locally finite quantifier  theory  $\bA$ over $\mS$ there is a global locally finite quantifier theory $\bA'$ such that $\bA = \bA'|_{\mS}$.
\\ (b) Any model of a global quantifier theory $\bA'$ induces a model of $\bA'|_{\mS}$ for any $\mS$.
}

  Suppose $\bA$ is a locally finite quantifier theory. Suppose $Z \in \mS$ is an infinite set. We say an ultrafilter $I$ of Boolean algebra $\bA_*(Z)$ is \la{perfect} if for any $z \in Z$ and $q \in \bA_*(Z)$ there is $d \in \bA^*(Z)$ such that $\forall z. q \vee \neg(q[d/z]) \in I$. Perfect ultrafilter plays the fundamental role in quantifier theories as that of ultrafilter in  Boolean algebras. The following theorem is a variant of ultrafilter theorem for Boolean algebras:

 \te{\label{te:hind} (G\"{o}del's Completeness Theorem for Quantifier Theories)  Suppose $\bA$ is a global  locally finite quantifier theory.  Suppose $X$ is an infinite set and $J$ is a consistent subset of $\bA_*(X)$.
\\ (a) There is an infinite set $X^+$ containing $X$ and a perfect ultrafilter $I$ of $\bA_*(X^+)$ containing $\kappa_{X*}(J)$, where $\kappa_X: X \ar X^+$ is the inclusion map.
\\ (b) The modification $\bA[X^+]/I$ of $\bA$ is a $2$-model of $\bA$ with $p\kappa_X = 1$ for any $p \in J$. }

Combining Theorem \ref{base}  and \ref{te:hind} we obtain:

\te{(G\"{o}del's Completeness Theorem for Quantifier Algebras) Suppose $\bA$ is a locally finite quantifier algebra over an infinite set $X$.
\\ (a) Suppose $J$ is a consistent subset of  $\bA_*(X)$. There is $2$-model $\bB$ of $\bA$ and a map $\sigma: X \ar \bB^*(\emptyset)$ such that $p\sigma = 1$ for any $p \in J$.
\\ (b) If $p, q \in \bA_*(X)$ and $p \ne q$ there is a $2$-model $\bB$ of $\bA$ and a map $\sigma: X \ar \bB^*(\emptyset)$ such that $p\sigma \ne q\sigma$.
}

The completeness theorems for quantifier theories are proved in Section 3 and 4. In Section 5 quantifier theories with equality are introduced, and ultraproducts of models of such theories are defined in Section 6. In Section 7 we define the notion of a polyadic theory. Since the category of locally finite polyadic algebras is equivalent to the category of locally finite quantifier algebras, the main theorems in \cite{halmos:1} for polyadic algebras can be easily derived from the completeness theorems for quantifier algebras.

In the second part of this paper we will study the free locally finite quantifier theory determined by a first-order language. Let $V = \{v_1, v_2, ...\}$ be a fixed countably infinite set of variables. For any integer $n \ge 0$ let $V_n = \{v_1, ..., v_n\}$. Let $\mL$ be a first-order language consisting of function and relation symbols. For every set $X$ let $\mL^*(X)$ be the set of $\mL$-terms over $X$. If $X$ is an infinite set containing $V$ let $\mL_*(X)$ be the set of $\mL$-formulas over $X$, modulo the relation $F \equiv G$ iff $\vdash F \Leftrightarrow G$ for any $F, G \in \mL_*(X)$.  If $X$ is any set let $X^+ = X \cup V$, and let $\mL_*(X) = \mL_*(X^+)_X$ be the subset of $\mL_*(X^+)$ determined by the $\mL$-formulas over $X^+$ with free variables in $X$. Then $\mL_*(X)$ is a Boolean (Lindenbaum) algebra  for any set $X$ (cf. \cite{bell:1}, p.191). Applying the (simultaneous) substitution theory of first-order logic (cf. \cite{bell:1}, p.65) we obtain a locally finite global quantifier theory $(\mL_*, \mL^*)$, called the \la{free global quantifier theory determined by the first-order language $\mL$}. Obviously $(\mL_*, \mL^*)$ has the following universal property:

 \te{ Suppose $V \in \mS$ and $\bA$ is a locally finite quantifier theory over $\mS$. Suppose $\mL$ is a first-order language, and  $\phi: \mL \ar \bA$ is function sending each $n$-ary function symbol to an element in $\bA^*(V)_{V_n}$ and each $n$-ary relation symbol to an element in $\bA_*(V)_{V_n}$. There is  a unique morphism $\Phi: (\mL_*, \mL^* )|_{\mS} \ar \bA$ such that $\Phi^*(f(v_1, ..., v_n)) = \phi(f)$ for each $n$-ary function symbol $f$, and  $\Phi_*(p(v_1, ..., v_n)) = \phi(p)$ for each $n$-ary relation symbol $p$.}

Let $\mL(\bA)$ be the first-order language with $\bA^*(V)_{V_n}$ as the set of $n$-ary function symbols and $\bA_*(V)_{V_n}$ as the set of $n$-ary relation symbols. Let $\phi: \mL \ar \bA$ be the map determined by the inclusion maps. Then the morphism $\Phi: (\mL(\bA)_*, \mL(\bA)^* )|_{\mS} \ar \bA$ given by the above theorem is surjective, which  determines an isomorphism from $\bA$ to the restriction $\bA'|_{\mS}$ of a quotient $\bA'$ of $(\mL(\bA)_*, \mL(\bA)^*)$. This yields another proof for the fundamental Theorem \ref{base}.

 \section{Properties of Binding Theories}

In this section we list some properties of a binding theory. For most of the statements, the proofs are straightforward and therefore will be omitted.

 \lm{\label{sup} Suppose $\bA$ is a (substitution or binding) theory. Assume $|X| > 1$. Suppose $x, y \in X$, $t \in \bA(X)$ and $U$ is a nonempty subset of $X$.
\\ (a) $U$ is  a support for $t$ if $t\sigma = t\tau$ for any two maps $\sigma, \tau: X \ar \bA(X)$ such that $\sigma|_U = \tau|_U$.
\\ (b) $t$ is independent of $x$ iff $t = t[y/x]$ for some $y \ne x$.
\\ (c)  $t$ is independent of $x$ iff $t = s[y/x]$ for some $s \in \bA(X)$ and $y \ne x$.
\\ (d) $U$ is a support for $t$ iff $t\gamma = t$ for a map $\gamma: X \ar X$ such that $\gamma(X) = U$ and $\gamma\gamma = \gamma$.
  \\ (e) The intersection of a finite collection of supports for $t$ is a support for $t$.
  \\ (f) If $\sigma: X \ar Y$ is injective (resp. bijective) then $\sigma*$$: \bA(X) \ar  \bA(Y)$ is injective (resp. bijective) and $\sigma*$$(\bA(X) ) = \bA(Y)_{\sigma(X)}$.
 }

 Let $\bA$ be a binding theory.

 \lm{Suppose $X \subseteq Y$. For any $x \in X$ and $p \in \bA_*(X)$ we have $(\forall x.p)\kappa_X = \forall x.(p\kappa_X)$ (note that $X \subseteq Y \subseteq \bA^*(Y)$). Thus if we identify $\bA(X)$ with $\bA(Y)_X$ via $\kappa_X*$ then $\forall x: \bA_*(X) \ar \bA_*(X)$ coincides with the restriction of $\forall x: \bA_*(Y) \ar \bA_*(Y)$ on $\bA_*(X)$, i.e., $\forall x = \forall x|_{\bA_*(X)}$.  }

\lm{\label{ind} Suppose $x \in X$ and $p \in \bA_*(X)$.
\\ (a) If $p$ has a support $U \subseteq X$, then $\forall x.p$ has a support $U \setminus \{x\}$. Thus $\forall x. p$ is independent of $x$.
\\ (b) $(\forall x. p)(\sigma[z/y]) = \forall y. (p(\sigma[z/y])^{y/x})$ for any map $\sigma: X \ar Y$, $z, y \in Y$ and $z \ne y$.
\\ (c) $(\forall x. p)\sigma = \forall y. (p\sigma^{y/x})$ for any map $\sigma: X \ar \bA^*(Y)$ and $y \in Y$ such that  $\sigma(z)$ is independent of $y$ for any $z$ in a support of $\forall x. p$ or $p$. }

\co{\label{sup}Suppose $\bA$ is a locally finite binding theory and  $Y$ is an infinite set. Then for any $x \in X$, $p \in \bA_*(X)$, and $\sigma: X \ar \bA^*(Y)$, we have $(\forall x. p)\sigma = \forall y. (p\sigma^{y/x})$ for some $y \in Y$ such that  $\sigma(z)$ is independent of $y$ for any $z$ in a support of $\forall x. p$ or $p$. }

\lm{\label{lm:q-less} Suppose $\bA$ is a quantifier theory. Suppose $p, q \in \bA_*(X)$.
\\ (a) If $p \leq q$ then $\forall x. p \leq \forall x.q$.
\\ (b) $\forall x.p \le p[z/x]$ for any $z \in \bA_*(X)$.
\\ (c) For any $p \in \bA_*(X)$, $\forall x. p$ is the largest element of $\{r \le p \ | \ r$ is independent of $x\}$.  }

\pf{(a) By Q1.
\\ (b) Since $\forall x. p \leq p$ by Q2 and $\forall x.p$ is independent of $x$ (Lemma \ref{ind}), we have $\forall x. p = (\forall x. p)[z/x] \leq p[z/x]$ for any $z \in X$.
\\ (c) If $r \le p$ and $r$ is independent of $x$ then $r = \forall x. r \le \forall x.  p$ by (a) and Q3. Thus $\forall x. p$ is the largest element of $\{r \le p \ | \ r$ is independent of $x\}$. }

\co{\label{co:q-less} A quantifier binding system on a substitution theory of Boolean algebras is unique if exists. }

\lm{\label{le:com} Suppose $\bA$ is a quantifier theory and $|X| > 2$. Write $\forall x_1...x_n.p$ for $\forall x_1.(... (\forall x_n.p)....)$. If $x, y \in X$ then $\forall y x. p = \forall x y. p$ for any $p \in \bA_*(X)$.
}

\pf{We may assume $x \ne y$. Assume $z$ is a variable which is different from $x, y$. Then $\forall x yx. p  = \forall x  y. ((\forall x .p)[z/x])) = \forall x .((\forall y  x. p)[z/x]) = (\forall y  x. p)[z/x] = \forall y .((\forall x .p)[z/x]) = \forall y x. p$ by Q3 and P5. Since $\forall x. p \leq p$ by Q2, we have $\forall x yx. p  \leq \forall x y. p$ by Lemma \ref{lm:q-less}. Thus $\forall y x. p \leq \forall x y. p$. Symmetrically we have $\forall x  y. p \leq \forall y x. p$.  Thus $\forall y x. p = \forall x y. p$.

}

\section{\label{section:function} Functional Theories}

Suppose $\mV$ is a variety of algebras.

Suppose $(B, M)$ is a pair consisting of a nonempty $\mV$-algebra $B$ and a nonempty set $M$. For any set $X$ denote by $M^X$ the set of maps from $X$ to $M$. Let $M^*(X) = M^{M^X}$ be the set of amps from $M^X$ to $M$, and let $B_*(X) = B^{M^X}$ be the set of maps from $M^X$ to $B$. We have a map $\epsilon: X \ar M^*(X)$ sending each $x \in X$ to the projection $\pi_x: M^X \ar M$ (with $\pi_x(\xi) = \xi(x)$) determined by $x$. We  shall identify $x$ with $\pi_x$. We identify $M$ with $M^*(\emptyset)$ and $B$ with $B_*(\emptyset)$.

If $p \in B_*(X)$ and $\sigma: X \ar M^*(Y)$ is a map we define $(p\sigma) \in B_*(Y)$ by $(p\sigma)(\xi) = p(\sigma\xi)$ for any $\xi: Y \ar M$, where $\sigma\xi: X  \ar M$ is defined by $(\sigma\xi)(x) = \sigma(x)(\xi)$. Similarly we define $a\sigma \in M^*(Y)$ for any $a \in M^*(X)$. Then $B_*(X)$ is a $\mV$-algebra pointwisely, and each map $\sigma_*: B_*(X) \ar B_*(Y)$ sending $p$ to $p\sigma$  is a homomorphism of $\mV$-algebras.

The structure $\bF(B, M) = (B_*, M^*)$ together with $a\sigma$ and $p\sigma$ defined above is a global substitution theory of $\mV$-algebras,  called the \la{$B$-valued functional substitution theory} determined by $M$.

Suppose $B$ is a Boolean algebra.  If $x \in X$ and $p \in B_*(X)$ define $\forall x.p \in B_*(X)$ such that $\forall x.p(\xi) = \bigwedge_{a \in M} p(\xi^{a/x})$ if the right side infimum exists for any $\xi: X \ar M$.  A subtheory $\bA$ of $\bF(B, M)|_{\mS}$ is called a \la{$B$-valued functional quantifier theory over $\mS$ } if $\forall x.p$ exists for any nonempty $X \in \mS$, $x \in X$, $p \in \bA_*(X)$, and $\forall x.p \in \bA_*(X)$. Then $\forall x$ is a well defined unary operation on $\bA_*(X)$. One can verify that these unary operations $\forall x$ have the following properties for any $p, q \in \bA_*(X)$:
\\ 1. $\forall x. (p \wedge q) = \forall x. p \wedge \forall x. q$,
\\ 2. $\forall x. p \leq  p$,
\\ 3. $\forall x. p = p$ if $p$ is independent of $x$.
\\ 4. $(\forall x.p)\sigma = \forall y.(p\sigma^{y/x})$ for any $p \in \bA_*(X)$, $\sigma: X \ar \bA^*(Y)$ and $y \in Y$ such that $\sigma(z)$ is independent of $y$ for any $z \in X$.
\\ Thus $\forall$ is a quantifier binding system on $\bA$. The pair $(\bA, \forall)$ is called a \la{$B$-valued functional quantifier theory over $\mS$}.

Suppose $\bA$ is a quantifier theory over $\mS$. A morphism of theories $\phi: \bA \ar \bF(B, M)$ is called a \la{quantifier morphism} if the image $\phi(\bA)$ of $\phi$ is a $B$-valued functional quantifier theory over $\mS$ and $\phi$ induces a morphism of quantifier theories from $\bA$ to $\phi(\bA)$.

\ex{Any quantifier model $\bA$ determines a morphism $\pi$ of quantifier theories $\bA \ar \bF(\bA_*(\emptyset), \bA^*(\emptyset))$ sending $a \in \bA^*(X)$ to $\pi(a) \in \bA^*(\emptyset)^{\bA^*(\emptyset)^X}$ with  $\pi(a)\sigma = a\sigma$ for any $\sigma: X \ar \bA^*(\emptyset)$, and sending $p \in \bA_*(X)$ to $\pi(p) \in \bA_*(\emptyset)^{\bA^*(\emptyset)^X}$ with  $\pi(p)\sigma = p\sigma$ for any $\sigma: X \ar \bA^*(\emptyset)$. Then $\bA$ is faithful iff the morphism $\pi$ is an embedding.
}

\te{ Suppose $B$ is a complete Boolean algebra and $M$ is a nonempty set. Then $\bF(B, M) = (B_*, M^*, \forall)$ is a global quantifier model. Any morphism from a quantifier theory $\bA$ to $\bF(B, M)$ determines a quantifier  model of $\bA$ with $\bA_*(\emptyset) = B$ and $\bA^*(\emptyset) = M$. Conversely, any quantifier model of $\bA$ arises in this way. In particular, if $B = 2$ then any nonempty set $M$ determines a global quantifier $2$-model $\bF(2_*, M^*)$.
}

\section{Cayley's Completeness Theorem}

\te{\label{cayley} Suppose $\bA$ is a locally finite quantifier theory. Suppose $Z \in \mS$ is an infinite set.
\\ (a) $\forall x. p = \bigwedge_{z \in Z} p[z/x]$ for any $x \in Z$ and $p \in \bA_*(Z)$.
\\ (b) $(\forall x. p)\sigma = \bigwedge_{z \in Z} p\sigma^{z/x}$ for any $x \in X$, $p \in \bA_*(X)$ and  $\sigma: X \ar \bA^*(Z)$.
\\ (c) $(\forall x. p)\sigma = \bigwedge_{d \in \bA^*(Z)} p\sigma^{d/x}$ for any $x \in X$, $p \in \bA_*(X)$, and  $\sigma: X \ar \bA^*(Z)$.
}

\pf{(a) We have $\forall x. p \leq p[z/x]$ for any $z \in Z$ by Lemma \ref{lm:q-less}. Thus $\forall x. p \leq \bigwedge_{z \in Z} p[z/x]$. Next assume $q \leq p[z/x]$ for every $z \in Z$. Since $Z$ is infinite and $\bA$ is locally finite, we can find $y, w \in Z$ such that $x, y, w$ are distinct and  $p$ and $q$ are independent of $y$ and $w$. Then $q \leq p[w/x]$ implies that $q[y/x] = q[y/x][x/w] \leq p[w/x][y/x][x/w] = p$. Since $q[y/x]$ is independent of $x$, we have  $q[y/x] = \forall x. (q[y/x]) \leq \forall x. p$. Then $q = q[y/x][x/y] \le (\forall x. p)[x/y] = \forall x. p$ as $p$, $q$ and $\forall x. p$ are independent of $y$. Hence $\forall x. p = \bigwedge_{z \in Z} p[z/x]$.
\\  (b) Since $Z$ is infinite and $\bA$ is locally finite, we have $(\forall x. p)\sigma = \forall_y (p\sigma^{y/x})$ for some $y \in Z$ such that  $\sigma(z)$ is independent of $y$ for any $z$ in a support of $p$ by Corollary \ref{sup}. Then \[(p\sigma^{y/x})[z/y]) = p\sigma^{z/x}.\] for any $z \in Z$.  Thus by (a) we have \[(\forall x. p)\sigma = \forall_y (p\sigma^{y/x}) = \bigwedge_{z \in Z} (p\sigma^{y/x})[z/y] = \bigwedge_{z\in Z} p\sigma^{z/x}.\]
\\  (c) Suppose $\{x_1, ..., x_n\}$ is a finite support for $p \in \bA_*(X)$. Since $Z$ is infinite and $\bA$ is locally finite, we can find $y \in Z$ such that $\sigma(x_1), ..., \sigma(x_n)$ are independent of $y$. Since $(\forall x. p)\sigma \le p\sigma^{y/x}$ by (b), we have $(\forall x. p)\sigma = (\forall x. p)\sigma[d/y]  \le (p\sigma^{y/x})[d/y] = p\sigma^{d/x}$ for any $d \in \bA^*(Z)$. Thus $(\forall x. p)\sigma \le \bigwedge_{d \in \bA(X)} p\sigma^{d/x}$. But by (b) we have $(\forall x. p)\sigma = \bigwedge_{z \in Z} p\sigma^{z/x} \ge \bigwedge_{d \in \bA(X)} p\sigma^{d/x}$. Thus $(\forall x. p)\sigma = \bigwedge_{d \in \bA(X)} p\sigma^{d/x}$.
}

\co{\label{co:cay} Suppose $Z$ is an infinite set, $p, q \in A_*(Z)$ and $x, y \in Z$. Suppose $p \le q[y/x]$ and $p, q$ are independent of $y$.
\\ (a) $p  \le q[z/x]$ for any $z \in Z$.
\\ (b) $p \le \forall x. q$.}

\pf{(a) Since $p \le q[y/x]$ and $p, q$ are independent of $y$, we have $p = p[z/y] \le q[y/x][z/y] = q[z/x]$ for any $z \in Z$. \\ (b) Since $\forall x. q = \bigwedge_{z \in Z} q[z/x]$ by Theorem \ref{cayley} (a),  we have $p \le \forall x. q$ by (a).
}

Suppose $\bA$ is a global theory. Any nonempty set $Z$ determines a modification $\bA[Z]$ of $\bA$ such that $\bA[Z]_*(\emptyset) = \bA_*(Z)$ and $\bA[Z]^*(\emptyset) = \bA^*(Z)$, with $t\sigma$ for any $t \in \bA(X)$ and $\sigma: X \ar \bA[Z]^*(\emptyset) = \bA(Z)$ as the same in $\bA$. $\bA[Z]$ is called the modification of $\bA$ by $Z$.

\te{\label{te:cay1} (Cayley's Completeness Theorem For Quantifier Theories) Suppose $\bA$ is a global locally finite quantifier theory. Suppose $Z$ is an infinite set. Then the modification $\bA[Z]$ of $\bA$ by $Z$ is a faithful model of $\bA$. }

\pf{Suppose $p, q \in \bA_*(X)$ and $p \ne q$. Suppose $U \subseteq X$ is a finite support for both $p, q$. Let $k: X \ar Z$ be a map such that $k|_U$ is injective. Then $k_*|_{\bA_*(X)_U}: \bA_*(X)_U \ar \bA_*(Z)$ is injective. Since $p, q \in \bA_*(X)_U$, we have $k_*(p) \ne k_*(p)$. The same analysis also apply to $a, b \in \bA^*(X)$ and $a \ne b$.  Hence the modification $\bA[Z]$ of $\bA$  is a faithful.  If $\bA$ is a quantifier theory then it satisfies M2 by Theorem \ref{cayley}, (c).  }

\te{\label{te:cay1} (Cayley's Completeness Theorem For Quantifier Algebras) Suppose $\bA$ is a locally finite quantifier algebra over an infinite set $X$. Then the modification $\bA[X]$ of $\bA$ by $X$ is a faithful model of $\bA$.  }

\te{\label{te:base-extension} (a) For any locally finite quantifier theory  $\bA$ over $\mS$ there is a global locally finite theory $\bA'$ such that $\bA = \bA'|_{\mS}$.
\\ (b) Any model of a global quantifier theory $\bA'$ induces a model of $\bA'|_{\mS}$ for any $\mS$.
}

\pf{(Sketch) Suppose $Z$ is an infinite set in $\mS$. The faithful model  $\bA[Z]$ of $\bA$ induces an embedding \[\pi: \bA \ar \bF(\bA_*(Z), \bA^*(Z)).\] Let $\bA'$ be the subtheory of $\bF(\bA_*(Z), \bA^*(Z))$ generated by the image $\pi(\bA)$ (i.e., the intersection of all subtheories containing $\pi(\bA)$). Then the theory $\bA$ is isomorphic to the restriction $\bA'|_{\mS}$. By Theorem \ref{cayley}  and Corollary \ref{co:q-less} there is a unique quantifier binding system $\forall$ on $\bA'$ such that $(\bA', \forall)$ is a locally finite quantifier theory. Then $\bA$ is isomorphic to the restriction $\bA'|_{\mS}$.
\\ (b) The assertion can be verified directly.
}

\section{\label{sec:com} G\"{o}del's Completeness Theorem}

Let $B$ be a nontrivial Boolean algebra. A \la{filter} of $B$ is a subset $I$ such that
for all $p, q$ in $I$ we have $p \wedge q$ in $I$ and $p \in I$, $r \geq p$ implies that $r \in I$. We say $I$ is \la{proper} if $I \ne B$, or equivalently, $0 \notin I$. An \la{ultrafilter} is a filter $I$ such that $p \in I$ iff $\neg p \notin I$ for any $p \in B$. A subset $J$ of $B$ is called \la{consistent} if it is contained in a proper filter. A subset $J$  is said to have the \la{finite meet property} (\la{f.m.p} for short) if whenever $p_1, ..., p_n \in J$ we have $p_1 \wedge ... \wedge p_n \ne 0$.

\lm{\label{rk:filter} (cf. \cite{bell:1}) If $J$ is any subset of $B$, let $\mF(J)$ be the set of elements $p \in B$ such that $p$ is larger than a finite intersection of elements in $J$. Then $\mF(J)$ is the filter generated by $J$. }

\lm{ (cf. \cite{bell:1}) 1. A subset $J$ is consistent iff $J$ has the finite meet property.
\\ 2. A subset is a maximal consistent subset iff it is an ultrafilter.
\\ 3. Any consistent set or proper filter is contained in an ultrafilter. More precisely, any proper filter is the intersection of all ultrafilters containing it.
}

Suppose $\bA$ is a global quantifier theory.
If $I$ is a filter of $\bA_*(\emptyset)$ and $\bA_*(\emptyset)/I$ is the quotient of the Boolean algebra $\bA_*(\emptyset)$ module $I$, we denote by $\bA/I$ the modification of $\bA$ with  $\bA_*(\emptyset) = \bA_*(\emptyset)/I$ and $\bA/I^*(\emptyset) = \bA^*(\emptyset)$.

Suppose  $\bA$ is a locally finite quantifier theory over $\mS$. Suppose $X \in \mS$ is an infinite set. We say an ultrafilter $I$ of $\bA_*(X)$ is \la{perfect} if for any $x \in X$ and $p \in \bA_*(X)$ there is $d \in \bA^*(X)$ such that $\forall x. p \vee \neg(p[d/x]) \in I$.

\lm{Suppose $Y$ is a countably infinite subset of $X$. An ultrafilter $I$ of $\bA_*(X)$ is perfect if for any $y \in Y$ and $p \in \bA_*(X)$ there is $d \in \bA^*(X)$ such that $\forall y. p \vee \neg(p[d/y]) \in I$.
}

\pf{By Corollary \ref{sup}, for any $x \in X$ and $p \in \bA_*(X)$  we have $\forall x. p = \forall y. (p[y/x])$ for some $y \in Y$ such that  $p$ is independent of $y$. By assumption  there is $d \in \bA^*(X)$ such that $\forall y. (p[y/x]) \vee \neg(p[y/x][d/y]) \in I$. Then $\forall x. p \vee \neg(p[d/x]) \in I$. Hence $I$ is prefect by definition.
}

The importance of the notion of ultrafilter lies in the following lemma:

\lm{\label{perfect1} Suppose $I$ is a perfect ultrafilter of $\bA_*(Z)$. The modification $\bA[Z]/I$ of $\bA$ is a $2$-model of $\bA$.
}

\pf{The condition M2 means that $(\forall x. p)\sigma \in I$ iff $p\sigma^{d/x} \in I$ for any $x \in X$, $p \in \bA_*(X)$, and $d \in \bA^*(Z)$. Suppose $(\forall x. p)\sigma \in I$. We have $(\forall x. p)\sigma \le p\sigma^{d/x}$ for any $d \in \bA^*(Z)$ by Theorem \ref{cayley}, (c).   Since $I$ is a filter of $\bA_*(Z)$, this implies that $p\sigma^{d/x} \in I$. Conversely, assume $(\forall x. p)\sigma \notin I$. Since  $\bA$ is locally finite we can find some $y \in Z$ such that $(\forall x. p)\sigma = \forall y. (p\sigma^{y/x})$ and  $\sigma(z)$ is independent of $y$ for any $z$ in a support of $p$ by Corollary \ref{sup}.  Hence $\forall y. (p\sigma^{y/x}) \notin I$. Since $I$ is perfect,  we have
$\forall y. (p\sigma^{y/x}) \vee \neg (p\sigma^{y/x}[d/y]) \in I$ for some $d \in \bA^*(Z)$. Thus $\forall y. (p\sigma^{y/x}) \notin I$ implies that $\neg(p\sigma^{y/x}[d/y]) = \neg (p\sigma^{d/x}) \in I$. Hence $p\sigma^{d/x} \notin I$. It follows that the modification $\bA[Z]/I$ of $\bA$ is a $2$-model of $\bA$. }

Suppose $\bA$ is a quantifier theory. Let $|\bA^*| = |\bA^*(X)_Z|$ and $|\bA_*| = |\bA_*(X)_Z|$ for any infinite set $X$ and any countably infinite subset $Z$ of $X$. Note that $\bA^*$ and $\bA_*$ are well-defined.

\lm{\label{lm:card1} Suppose $\bA$ is a locally finite quantifier theory and $X$ is an infinite set.
\\ (a) $|\bA_*(X)| = |X| + |\bA_*| = |X|\cdot |\bA_*| = Max(|X|, |\bA_*|)$ (cf. \cite{end:1}, p.164). In particular, $|\A_*|$ is infinite, and if $|\bA_*|$ is countably infinite then $|\bA_*(X)| = |X|$.
\\ (b) $|\bA^*(X)| = |X| + |\bA^*| = |X|\cdot |\bA^*| = Max(|X|, |\bA^*|)$ . In particular, if $|\bA^*|$ is countably infinite then $|\bA^*(X)| = |X|$.
}

\pf{(a) Suppose $Z = \{z_1, z_2, ...\}$ is a countably infinite subset of $X$. For any $x_1, ..., x_m \in X$ let $[x_1, ..., x_m]: Z \ar X$ be the map sending $z_i$ to $x_i$ ($i = 1, ..., m$) and any other $z \in Z$ to itself. Since $\bA$ is nontrivial and $Z$ is infinite we can find a non-closed  element $q \in \bA_*(Z)$. Suppose $\{z_1, z_2, ..., z_n\}$ ($n > 0$) is  a minimal support for $q$. Let $\pi: X \times \{1, ..., n\} \ar X$ be a bijective map. Define a map $\delta: X \ar \bA_*(X)$ sending $x$ to $q[\pi(x, 1), ... \pi(x, n)]$. Since $\{\pi(x, 1), ... \pi(x, n)\}$ is a set of $n$ distinct variables, $\delta(x)$ is non-closed with a support $U_x = \{\pi(x, 1), ... \pi(x, n)\}$. If $x \ne y$ then $\delta(x)$ and $\delta(y)$ are non-closed elements with disjoint minimal supports $U_x \ne U_y$. Thus $\delta(x) \ne \delta(y)$. Hence $\delta$ is injective. So $ |\bA_*(X)| \ge |X|$. Since $Z \subseteq X$ we have $\bA_*(X) \supseteq \bA_*(Z)$. Thus  $|\bA_*(X)| \ge |\bA_*(Z)| = |\bA_*|$.  It follows that $|\bA_*(X)| \ge |X| + |\bA_*| = |X| \cdot |\bA_*| = Max(|X|, |\bA_*|)$. For any element $p \in \bA_*(X)$ we can find an element $p' \in \bA_*(Z)$ and a sequence $\{x_1, ..., x_n\} \subset X$ such that $p = p'[x_1, ..., x_n]$. The map sending each $p \in \bA_*(X)$ to $<$$p', x_1, ..., x_n$$>$ is an injective map from $\bA_*(X)$ to the set of finite sequences of elements in $\bA_*(Z) \cup X$. Thus $|\bA_*(X)| \le |\bA_*(Z)| + |X|$. It follows that $|\bA_*(X)| = |X| + |\bA_*| = |X|\cdot |\bA_*| = Max(|X|, |\bA_*|)$.
\\ (b) The proof is similar. }

\te{\label{te:god} (G\"{o}del's Completeness Theorem for Quantifier Theories)  Suppose $\bA$ is a global  locally finite quantifier theory.  Suppose $X$ is an infinite set and $J$ is a consistent subset of $\bA_*(X)$.
\\ (a) There is an infinite set $X^+$ containing $X$ and a perfect ultrafilter $I$ of $\bA_*(X^+)$ containing $\kappa_{X*}(J)$, where $\kappa_X: X \ar X^+$ is the inclusion map.
\\ (b) The modification $\bA[X^+]/I$ of $\bA$ is a $2$-model of $\bA$ with $p\kappa_X = 1$ for any $p \in J$.}

\pf{ We may assume that $\bA$ is nontrivial.
\\ (a) Let $\lambda = |\bA_*(X)|$. Since $\bA$ is nontrivial locally finite and $X$ is infinite, we have $\lambda = |\bA_*(X)| = |X| \cdot |\bA_*| = Max(|X|, |\bA_*|)$ by Lemma \ref{lm:card1}; thus $\lambda \ge |X|$ and $\lambda \ge |\bA_*|$. Let $X^+$ be a set containing $X$ such that $X^+\setminus X$ has cardinality $\lambda$; variables in  $X^+\setminus X$ are called \la{new variables}. Then $|X^+| = \lambda$, and $\bA_*(X^+)$ is infinite by Lemma \ref{lm:card1}.  Let $Y$ be a countably infinite subset of $X$. Then $|Y \times \bA_*(X^+)| = |Y|\cdot |\bA_*(X^+)| = Max(|Y|, |\bA_*(X^+)|) =  |\bA_*(X^+)| = |X^+| \cdot |\bA_*| = |X^+| \cdot |\bA_*| =  \lambda \cdot |\bA_*| = \lambda$ again by Lemma \ref{lm:card1}. We fix a well-ordering \[<y_{\alpha}, p_{\alpha}>_{\alpha < \lambda}\] of the set $Y \times \bA_*(X^+)$. For $\alpha < \lambda$ let \[\theta_{\alpha} = \forall y_{\alpha}. p_{\alpha} \vee \neg (p_{\alpha}[z_{\alpha}/y_{\alpha}]),\] where $z_{\alpha}$ is the first new variable such that $p_{\alpha}$ and $\theta_{\beta}$ are independent of $z_{\alpha}$ for any $\beta < \alpha$. (This excludes at most $|\alpha|$ new variables, so there are some left.) Let \[\Theta = \{\theta_{\alpha} | \alpha < \lambda\}, \ \ \ \Gamma = \kappa_{X*}(J) \cup \Theta.\]  Since $J$ is consistent,  it has f.m.p. Since $\kappa_{X*}$ is an injective homomorphism of Boolean algebras,  $\kappa_{X*}(J)$ has also f.m.p. Thus $\kappa_{X*}(J)$ is consistent. Also $X$ is a support for any member of $\kappa_{X*}(J)$. Thus any member of $\kappa_{X*}(J)$ is independent of any new variable. We show that $\Gamma$ is consistent. Assume this  is not true. Then there is a finite intersection $p \ne 0$ of members of $\kappa_{X*}(J)$, and $\alpha_1 < ... < \alpha_m < \alpha < \lambda$ such that \[p \wedge \theta_{\alpha_1} \wedge ...\wedge \theta_{\alpha_m} \wedge \theta_{\alpha} = 0.\] Take the least such $\alpha$. Let \[q = p \wedge \theta_{\alpha_1} \wedge ... \theta_{\alpha_m}.\] Then \[q \ne 0, \ \ \ q \wedge \theta_{\alpha} = 0.\]  Since \[\theta_{\alpha} = \forall y_{\alpha}. p_{\alpha} \vee \neg (p_{\alpha}[z_{\alpha}/x_{\alpha}]),\] we have \[q \wedge(\forall y_{\alpha}. p_{\alpha} \vee \neg (p_{\alpha}[z_{\alpha}/x_{\alpha}]) = 0 .\] Thus \[(q \wedge \forall y_{\alpha}. p_{\alpha})  \vee (q \wedge \neg (p_{\alpha}[z_{\alpha}/x_{\alpha}])) = 0.\] This implies \[q \wedge \forall y_{\alpha}. p_{\alpha} = 0, \ \ \ q \wedge \neg (p_{\alpha}[z_{\alpha}/x_{\alpha}])= 0.\]     Hence we have \[q \le \neg(\forall y_{\alpha}. p_{\alpha}), \ \ \  q \le  p_{\alpha}[z_{\alpha}/x_{\alpha}].\]  Since $q, p_{\alpha}$ are independent of the new variable $z_{\alpha}$, applying Corollary \ref{co:cay} to $q \leq p_{\alpha}[z_{\alpha}/x_{\alpha}]$ we conclude that \[q \le \forall y_{\alpha}. p_{\alpha}.\] Thus \[q \leq  \forall y_{\alpha}. p_{\alpha} \wedge \neg(\forall y_{\alpha}. p_{\alpha}) =  0.\]  We obtain $q = 0$, which contradicts to the assumption that $q \ne 0$. This shows that $\Gamma$ has f.m.p. Hence $\Gamma$ is consistent. Let $I$ be an ultrafilter in $\bA_*(X^+)$ containing $\Gamma$. Then $I$ is a perfect ultrafilter in $\bA_*(X^+)$. Also $\kappa_{X*}(J) \subset \Gamma \subset I$.
\\ (b) By Lemma \ref{perfect1} we conclude  that $\bA[X^+]/I$ of $\bA$ is a $2$-model of $\bA$, and  $p\kappa_X = \kappa_{X*}(p) \in I$ for any $p \in J$. Thus $p\kappa_X = 1$.
}

Combining Theorem \ref{te:god}  and \ref{te:base-extension} we obtain:

\te{(G\"{o}del's Completeness Theorem for Quantifier Algebras) Suppose $\bA$ is a locally finite quantifier algebra over an infinite set $X$.
\\ (a) Suppose $J$ is a consistent subset of  $\bA_*(X)$. There is $2$-model $\bB$ of $\bA$ and a map $\sigma: X \ar \bB^*(\emptyset)$ such that $p\sigma = 1$ for any $p \in J$.
\\ (b) If $p, q \in \bA_*(X)$ and $p \ne q$ there is a $2$-model $\bB$ of $\bA$ and a map $\sigma: X \ar \bB^*(\emptyset)$ such that $p\sigma \ne q\sigma$.
}

\pf{(b) Since $p \ne q$, either $p \nleq q$ or $q \nleq p$. Assume the first case holds. Then $p \wedge (\neg q) \ne 0$. By (a) we can find a $2$-model $\bB$ of $\bA$ and a map $\sigma: X \ar \bB^*(\emptyset)$ such that $(p \wedge \neg q) \sigma = 1$. Then $(p \wedge \neg q) \sigma = (p \sigma) \wedge (\neg q) \sigma = 1$. Hence $p \sigma = 1$ and $(\neg q) \sigma = \neg (q \sigma) =  1$. Thus $q\sigma = 0$. It follows that $p\sigma \ne q\sigma$. }

 \section{Quantifier Theories with Equality}

 Let $\bA$ be a quantifier theory. An \la{equality $e$ } of $\bA$ consists of an element $e(a, b) \in \bA_*(X)$ for any $a, b \in \bA^*(X)$ such that the following condition is satisfied:
\\ E1. $e(a, b)\sigma = e(a\sigma, b\sigma)$ for any $a, b \in \bA^*(X)$ and $\sigma: X \ar \bA^*(Y)$.
\\ E2. $e(a, a) = 1$ for any $a \in \bA^*(X)$;
\\ E3. $p \wedge e(x, y) \le p[x/y]$ for any $x, y \in X$ and $p \in \bA_*(X)$.
\\A \la{quantifier theory with equality} is a quantifier theory together with an equality $e$ of $\bA$.
\\ A \la{normal quantifier model} is a quantifier model with equality such that the following condition is satisfied:
   \\ M3. For any two elements $a, b \in \bA^*(\emptyset)$ we have $e(a, a) = 1$ and  $e(a, b) = 0$ if $a \ne b$.

 Suppose $\bA$ and $\bB$ are quantifier theories with equality. By a \la{morphism $\phi$ of quantifier theories with equality from $\bA$ to $\bB$}  we mean a morphism of binding theories such that the following condition is satisfied:
\\ N5  $\phi_*(e(a, b)) = e(\phi^*(a), \phi^*(b))$.

   \lm{Suppose $e$ is an equality of a quantifier theory $\bA$.
   \\ (a) $p \wedge e(x, y) = p[x/y] \wedge e(x, y)$.
   \\ (b) $e(x, y)$ is the smallest element $p$ of $\bA_*(X)$ such that $p[y/x] = 1$.
   }

   \pf{(a) By E3 we have $p[y/x] \wedge (\neg p) \wedge e(x, y) \le p[y/x] \wedge (\neg p)[x/y] = (p \wedge \neg p)[y/x] = 0[y/x] = 0$. Thus $p[y/x] \wedge  e(x, y) \le p$. So $p[y/x] \wedge  e(x, y) \le p \wedge e(x, y)$. But E3 implies that $p \wedge e(x, y) \le p[x/y] \wedge e(x, y)$. Thus $p \wedge e(x, y) = p[x/y] \wedge e(x, y)$.
   \\ (b) We have $e(x, y)[y/x] = e(y, y) = 1$. Next if $p[y/x] = 1$ then by (a) we have $p \wedge e(x, y) = 1 \wedge e(x, y)$. Thus $e(x, y) \le p$.

   }

   \co{An equality of a quantifier theory is unique if exists. }

Suppose $\bA$ is a locally finite quantifier $2$-model with equality. Denote by the equivalence relation $\theta$ on $\bA^*(\emptyset)$ such that $a\theta b$ iff $e(a, b) = 1$. Then $\theta$ is a congruence on $\bA^*(\emptyset)$ in the sense that $p\sigma = p\tau$ and $a\sigma \theta a\tau$ for any $a \in \bA^*(X)$, $p \in \bA_*(X)$,  and $\sigma, \tau: X \ar \bA^*(\emptyset)$ such that $\sigma(x) \theta \tau(x)$ for any $x \in X$. Let $\bA^*(\emptyset)/\theta$ be the quotient of $\bA^*(\emptyset)$ by $\theta$. Let  $\bA/\theta$ be the modification of $\bA$ with $(\bA/\theta)_*(\emptyset) = \bA_*(\emptyset)$ and $(\bA/\theta)^*(\emptyset) = \bA^*(\emptyset)/\theta$

\lm{\label{lm:normal-equ} The modification $\bA/\theta$ of $\bA$ is a normal quantifier model. }

It follows that Theorem \ref{te:god} also applies to locally finite quantifier theory with equality:

\te{\label{te:god} (G\"{o}del's Completeness Theorem for Quantifier Theories with Equality)  Suppose $\bA$ is a global locally finite quantifier theory with equality.  Suppose $X$ is an infinite set and $J$ is a consistent subset of $\bA_*(X)$. There is an infinite set $X^+$ containing $X$ and a perfect ultrafilter $I$ of $\bA_*(X^+)$ containing $\kappa_{X*}(J)$, where $\kappa_X: X \ar X^+$ is the inclusion map.
 The modification $(\bA[X^+]/I)/\theta$ of $\bA$ is a normal $2$-model of $\bA$ with $p\kappa_X = 1$ for any $p \in J$. }

 \te{(G\"{o}del's Completeness Theorem for Quantifier Algebras with Equality) Suppose $\bA$ is a locally finite quantifier algebra with equality over an infinite set $X$. Suppose $J$ is a proper filter of  $\bA_*(X)$. There is normal $2$-model $\bB$ of $\bA$ and a map $\sigma: X \ar \bB^*(\emptyset)$ such that $p\sigma = 1$ for any $p \in J$.
}

\section{Ultraproducts of Models}

Let $\mI$ be a nonempty index set. Let $\bA$ be a  locally finite quantifier theory. For each $i \in \mI$ let $\bB_i$ be a $2$-model of $\bA$. Let $\prod_{i\in \mI} \bB_i^*(\emptyset)$ and $\prod_{i \in \mI} 2$ be the Cartesian products. Denote by $\bB$ the modification of $\bA$ with $\bB_*(\emptyset) = \prod_{i \in \mI} 2$ and $\bB^*(\emptyset) = \prod_{i\in \mI} \bB_i^*(\emptyset)$, such that $(a\sigma)(i) = a\sigma_i$  and $(p\sigma)(i) = p\sigma_i$ for any $a \in \bA^*(X)$, $p \in \bA_*(X)$, $\sigma: X \ar \bB^*(\emptyset)$, where $\sigma_i: X \ar \bB_i^*(\emptyset)$ is defined by $\sigma_i(x) = \sigma(x)(i)$ for any $x \in X$.

\te{ (a) $\bB$ is a model of $\bA$.
\\ (b) If  $I$ is an ultrafilter of $\bB_*(\emptyset) = \prod_{i \in \mI} 2$ then $\bB/I$ is a $2$-model of $\bA$. }

\pf{(a) We prove that for any $x \in X$, $p \in \bB_*(X) = \bA_*(X)$, and $\sigma: X \ar \bB^*(\emptyset)$ we have $(\forall x.p)\sigma = \bigwedge_{d \in \bB^*(\emptyset)} p\sigma^{d/x}$. For any $i \in \mI$ we have $((\forall x.p)\sigma)(i) = (\forall x.p)\sigma_i \le p\sigma_i^{d(x)/x} = (p\sigma^{d/x})(i)$ for any $d \in \bB^*(\emptyset)$, thus  $(\forall x.p)\sigma \le \bigwedge_{d \in \bB^*(\emptyset)} p\sigma^{d/x}$. Conversely, suppose $i \in \mI$ such that $p\sigma^{d/x}(i) = p\sigma_i^{d(i)/x} = 1$ for any $d \in \bB^*(\emptyset)$. Since $d(i)$ could be any element in $\bB_i^*(\emptyset)$, we have $p\sigma_i^{d'/x} = 1$ for any $d' \in \bB_i^*(\emptyset)$. Then $(\forall x.p\sigma_i) = \bigwedge_{d' \in \bB_i^*(\emptyset)}p\sigma_i^{d'/x} = \bigwedge_{d' \in \bB_i^*(\emptyset)} 1 = 1$ as $\bB_i$ is a model of $\bA$. Thus $(\forall x.p)\sigma \ge \bigwedge_{d \in \bB^*(\emptyset)} p\sigma^{d/x}$. Hence $(\forall x.p)\sigma = \bigwedge_{d \in \bB^*(\emptyset)} p\sigma^{d/x}$.
\\ (b)  We  prove that for any $x \in X$, $p \in  \bB_*(X)$, and $\sigma: X \ar \bB^*(\emptyset)$ we have $(\forall x.p)\sigma = \bigwedge_{d \in \bB^*(\emptyset)} p\sigma^{d/x}$ in $\bB_*(\emptyset)/I$, i.e., $(\forall x.p)\sigma \in I$ iff $p\sigma^{d/x} \in I$ for any $d \in \bB^*(\emptyset)$.  First assume $(\forall x.p)\sigma \in I$. Since by (a) we have $(\forall x.p)\sigma = \bigwedge_{d \in \bB^*(\emptyset)} p\sigma^{d/x}$, thus $(\forall x.p)\sigma \le  p\sigma^{d/x}$ for any $d \in  \bB^*(\emptyset)$, so $p\sigma^{d/x} \in I$ for any $d \in  \bB^*(\emptyset)$ as $I$ is a filter. Next assume $(\forall x.p)\sigma  \notin I$. We have to find $d': X \ar \bB_*(\emptyset)$ such that $p\sigma^{d'/x} \notin I$. Since $I$ is an ultrafilter, we have $(\neg \forall x.p)\sigma  \in I$. Suppose $(\neg \forall x.p)\sigma(i) = 1$. Then $(\neg \forall x.p)\sigma_i = 1$. So $(\forall x.p)\sigma_i = 0$. Since $\bB_i$ is a model, there is $d_i \in \bB_i^*(\emptyset)$ such that $p\sigma^{d_i/x} = 0$. Let $d': X \ar \bB_*(\emptyset)$ be any map such that $d(i) = a_i$ for any $i$ with $(\neg \forall x.p)\sigma(i) = 1$. Then $(\neg \forall x.p)\sigma(i) = 1$ implies that $ p\sigma^{d'/x}(i) = 0$, i.e., $\neg p\sigma^{d'/x}(i) = 1$. Thus $(\neg \forall x.p)\sigma \le \neg p\sigma^{d'/x}$. Hence $\neg p\sigma^{d'/x} \in I$ as $I$ is a filter. Since $I$ is an ultrafilter, we have $p\sigma^{d'/x} \notin I$. This finish the proof.

}

Next assume $\bA$ is a quantifier theory with equality and each $\bB_i$ is a normal $2$-model of $\bA$. Suppose $I$ is an ultrafilter of $\bB_*(\emptyset) = \prod_{i \in \mI} 2$.  Let $\theta$ be the equivalence relation on $(\bB/I)^*(\emptyset) = \prod_{i\in \mI} \bB_i^*(\emptyset)$ such that $a\theta b$ iff $e(a, b) = 1$ in $(\bB/I)_*(\emptyset) = (\prod_{i \in \mI} 2)/I$ for any $a, b \in \prod_{i\in \mI} \bB_i^*(\emptyset)$.

Applying Lemma \ref{lm:normal-equ} we obtain the following theorem, which implies $\L$o\'{s}'s Ultraproduct Theorem in model theory (see \cite{bell:1} p.180):

\te{ Under the above assumptions $(\bB/I)/\theta$ is a normal $2$-model of $\bA$. }

\section{Polyadic Theories}

 A \la{polyadic theory (over $\mS$}) consists of a substitution theory $\bA$ of Boolean algebras (over $\mS$) together with a map $\forall_U: \bA_*(X) \ar \bA_*(X)$ for any nonempty set $X \in \mS$, and $U \subseteq X$ such that for any $p, q \in \bA_*(X)$:
\\ 1. $\forall_{ \emptyset} p = p$.
\\ 2. $\forall_{ U \cup V} p = \forall_U \forall_V p$ for any  $U, V \subseteq X$.
\\ 3. $( \forall_U p)\sigma = \forall_V ( p\sigma^{\pi})$ for any map $\sigma: X \ar \bA^*(Y)$, $U \subseteq X$, $V \subseteq Y$,  and any injective map $\pi: U \ar V$ such that $\sigma(x)$ is independent of  $V$ for any $x \in X$.
\\ 4. $\forall_U (p \wedge q) = \forall_U p \wedge \forall_U q $.
\\ 5. $\forall_U p \leq p$.
\\ 6. $\forall_U p = p$ if $p$ is independent of $U$.
\\ If  $\mS = \{X\}$ then $\bA$ is a \la{polyadic algebra} over $X$ in the sense of Halmos \cite{halmos:1}.

A \la{polyadic model} is a polyadic theory $\bA$ over $\mS$ with $\emptyset \in \mS$ satisfying the following conditions:
 \\ M1. $\bA^*(\emptyset)$ is nonempty and $\bA_*(\emptyset)$ is a nontrivial Boolean algebra.
  \\ M2. For any $U \subseteq X$, $p \in \bA_*(X)$, and $\sigma: X \ar \bA^*(\emptyset)$ we have $(\forall_U.p) = \bigwedge \{ p\tau \ | \ \tau|_{X\setminus U} = \sigma|_{X\setminus U}$\}, where $\tau: X \ar \bA^*(\emptyset)$ is a map.
   \\ We say $\bA$ is a \la{polyadic $2$-model} if  $\bA_*(\emptyset) = 2 = \{0, 1\}$.
 \\ Suppose $\bA$ is a polyadic theory over $\mS$. A \la{modification} (resp. model) of $\bA$ is a polyadic theory (resp. polyadic model) $\bB$ over $\mS \cup \{\emptyset\}$ such that $\bA|_{\mS\setminus \{\emptyset\}} = \bB|_{\mS\setminus \{\emptyset\}}$.

Any polyadic quantifier theory induces a quantifier theory with $\forall x = \forall\{x\}$. Conversely, any locally finite quantifier theory  determines a locally finite polyadic theory with $\forall_U p = \forall x_1...x_n. p$ for any  set $U \subseteq X$, where $\{x_1, ... , x_n\} \subseteq U$ is any finite support for $p$ (cf. Lemma \ref{le:com}). Hence the notion of locally finite polyadic theory is equivalent to that of locally finite quantifier theory (cf. \cite{pinter:firstorder}).  For other approaches to the theory of polyadic algebras see  \cite{cir:0} - \cite{Dai:3}, \cite{Gai:1} - \cite{leb:2} and \cite{pinter:firstorder} - \cite{plotkin:1}.

Note that Theorem \ref{te:base-extension} also applies to locally finite  polyadic theories:

 \te{\label{te:base-extension1} (a) For any locally finite  polyadic theory  $\bA$ over $\mS$ there is a global theory $\bA'$ such that $\bA = \bA'|_{\mS}$.
\\ (b) Any model of a global polyadic theory $\bA'$ induces a model of $\bA'|_{\mS}$ for any $\mS$.
}

\end{document}